\documentclass[12pt]{iopart}
\newcommand{\mb}[1]{\mbox{\boldmath $#1$}}
\newcommand{\npb}{$\{\mb{k},\mb{\ell},\mb{m},\mb{\bar{m}}\}\;$}
\newcommand{\al}{\mbox{$\not\!\alpha$}}
\newcommand{\bb}{\mbox{$\not\!\beta$}}
\newcommand{\plusi}{\mbox{$\,+\,i$}}

\begin{document}

\jl{6}
    
\title[Consequences of a Killing symmetry in spacetime's local structure]
{Consequences of a Killing symmetry in spacetime's local structure}

\author{Francesc Fayos\dag$\S$\ and Carlos F. Sopuerta\ddag
\footnote[3]{Also at the Laboratori de F\'{\i}sica Matem\`atica,
Societat Catalana de F\'{\i}sica, I.E.C., Barcelona, Spain}}

\address{\dag\ Departament de F\'{\i}sica Aplicada, UPC,
E-08028 Barcelona, Spain}
 
\address{\ddag\ Institute of Cosmology and Gravitation, Mercantile 
House, Hampshire Terrace, PO1 2EG Portsmouth, United Kingdom}

~

\address{E-mail: {\tt labfm@ffn.ub.es}\,, 
{\tt carlos.sopuerta@port.ac.uk}}

\begin{abstract}
In this paper we discuss the consequences of a Killing symmetry on the
local geometrical structure of four-dimensional spacetimes.  
We have adopted the point of view introduced in recent works where the 
exterior derivative of the Killing plays a fundamental role.  Then, we 
study some issues related with this approach and clarify why in many 
circumstances its use has advantages with respect to other approaches.  
We also extend the formalism developed in the case of vacuum spacetimes 
to the general case of an arbitrary energy-momentum content.   Finally, 
we illustrate our framework with the case of spacetimes with a 
gravitating electromagnetic field. 
\end{abstract}


% Uncomment for Submitted to journal title message
%\submitted

% Comment out if separate title page not required
%\maketitle

%%%%%%%%%%%%%%%%%%%%%%%%%%%%%%%%%%%%%%%%%%%%%%%%%%%%%%%%%%%%%%%%%%%%%%%%%%
%
%                INTRODUCTION (SECTION ONE)
%
%%%%%%%%%%%%%%%%%%%%%%%%%%%%%%%%%%%%%%%%%%%%%%%%%%%%%%%%%%%%%%%%%%%%%%%%%%

\section{Introduction}
The imposition of symmetries is and has been one the most successful ways of 
simplifying the theory in order to deal with the essence of physical problems.  
In the case of General Relativity they have been extensively used to 
simplify Einstein's field equations, and in this way to obtain exact 
solutions describing the gravitational field of the idealized situation.
Of course, due to the non-linear character of the field equations,
assuming many symmetries can lead to solutions which may not capture the
actual behaviour of the physical system they are meant to represent. 
Then, it is possible that {\em deviations} from the idealized situation
behave in a very different way.  Nevertheless, we know of many systems whose
dynamics is driven by the gravitational interaction and where there is a
symmetry that is present in the spacetime region of interest.  Then,
it is important to know how the imposition of a symmetry affects to
the structure of the spacetime and which can be the consequences for
a specific problem.

In this paper we will consider spacetimes with only one Killing symmetry,
which can describe many physical situations of interest.   Examples
of present interest are provided by physical systems in quasi-equilibrium
configurations,  like neutron-star binaries around the innermost
stable circular orbit (ISCO), which are of great interest in the study of 
gravitational wave emission by binary systems.  In this context, methods 
to construct such configurations in order to locate the ISCO rely on
the assumption of the existence of an approximate Killing vector~\cite{BCSST}.
Other problems of interest in which a symmetry is present are those involving 
axisymmetric configurations, stationary systems, etc.

There are many papers in the literature dealing with spacetimes possesing
Killing vector fields (KVFs hereafter).  Most of these papers are 
devoted to the search of exact solutions of Einstein's field equations,
and KVFs were considered as a way of simplifying
the problem (see~\cite{KSHM}).   Then, the main aim was to develop and apply
techniques to solve the Einstein equations when a group of Killing symmetries 
was imposed.  Despite the huge amount of work done there are few studies
that concentrate only on the study of spacetimes containing just one KVF.  
Then, few is known about the geometrical properties of 
a Killing symmetry in connection with the spacetime geometry.  Examples
of works where formalisms for one KVF are introduced are the works by 
Collinson and French~\cite{COFR} (for conformal KVFs) and by Perj\'es~\cite{PERJ}.
In the first work the equations (including the Killing equations) are
written using the Newman-Penrose formalism~\cite{NEPE}, which leads to 
simplifications when the NP basis is adapted to the algebraic structure
of the spacetime.   However, it has the limitation that it does not 
introduce any quantity adapted to the (conformal) KVF.  Then, by one hand
it can be used for groups of (conformal) KVFs but, on the other hand, it
cannot deal with the particular characteristics of the symmetry.
In contrast, the work by Perj\'es combines in a powerful way
the spinor formalism (the NP formalism is a byproduct of it) and quantities
describing the structure of the symmetry.  However, since it uses
quantities related to the Ernst potential, the generalization to general
spacetimes with one KVF is not obvious at all.

Recently, we introduced~\cite{FASO1,FASO2} a new approach to the study of 
vacuum spacetimes with one KVF which is adapted to the structure of the 
Killing symmetry and which can be easily generalized to any matter content 
(this is one of the subjects of the present paper).  As we showed in the
previous works~\cite{FASO1,FASO2}, this approach naturally leads to
a classification~\cite{FASO2} of the spacetimes with at least one symmetry,
which is based on the properties of the KVF and on the relations of its 
algebraic structure with that of the spacetime (the algebraic structure
of the Weyl tensor).  Moreover, in~\cite{FASO1,FASO2} we showed the potential 
of this approach by proving a number of new results in the vacuum case 
(see~\cite{STEELE1} for an error in one of them).  
Generalizations of this formalism have recently appeared for the case of 
homotheties~\cite{STEELE2} and conformal KVFs~\cite{LUDWIG}.

The approach introduced in~\cite{FASO1,FASO2} considers the exterior derivative 
of the Killing, which we will call the {\em Papapetrou field} after its 
introduction in~\cite{PAPA} (sometimes also called the Killing 2-form 
or the Killing bivector~\cite{DEB1,DEB2}), and its algebraic structure,
as main objects of the study.  The Papapetrou field has been used before
in the search of exact solutions (see, e.g.,~\cite{GUYS}) and in the study of 
external magnetic fields in black holes~\cite{WALD}.  As we discussed
in detail in~\cite{FASO1,FASO2}, one main advantage of this approach is
that it provides a framework in which one can establish, in a natural way, 
connections between the particular characteristics of the symmetry 
(the structure of the Papapetrou field) and the 
algebraic structure of the spacetime (the Petrov type of the Weyl 
tensor~\cite{PETROV}) which, in particular, leads to a detailed 
classifications of the spacetimes with an isometry~\cite{FASO2}.  
Moreover, with the introduction of the Papapetrou field as a main 
quantity we constructed a new formalism for the study of vacuum 
spacetimes with an KVF~\cite{FASO2}.  It is an extension of the 
Newman-Penrose formalism~\cite{NEPE}, where new variables 
associated with the KVF and its Papapetrou field are introduced.  
Within this formalisms, the integrability conditions for the components
of the KVF are analyzed from a new perspective.  Using this
new point of view it was shown in~\cite{FASO2} that from the integrability 
conditions one gets expressions for all the Weyl tensor components in 
terms of the other variables, reducing in this way the number of
equations to be considered (we avoid the use of the second Bianchi 
identities).

In this paper we analyze in detail the structure of this approach 
and the reasons why it has advantages over other approaches, and we extend
the formalism to spacetimes with a KVF and with a generic energy-momentum 
content.  The plan of the paper is the following:  In section~\ref{alge} 
we present the main ideas and justify why the adoption of the Papapetrou 
field as a main quantity means to have advantages when dealing with 
spacetimes with a KVF.  At the same time we study in detail, from a new
perspective, the integrability conditions for the components of the KVF.
This part will also involve an explanation of previous 
results~\cite{FASO1,FASO2} 
in a broader context. In section~\ref{papi} we study the integrability 
conditions for the Papapetrou field, which satisfies Maxwell equations.  
In section~\ref{genf} we extend the framework introduced in~\cite{FASO2} 
for vacuum to general spacetimes.  For the sake of clarity we will give the 
equations that come out from this 
formalism in the~\ref{appa}, where they have been expressed using the 
Newman-Penrose formalism~\cite{NEPE}.  In section~\ref{eima} we illustrate 
these developments with the case of spacetimes with a gravitating 
electromagnetic field, and in particular, with the case of the Kerr-Newman
metric.  We finish with some remarks and conclusions in section~\ref{reco}. 
Throughout this paper we adopt units in which $c = 8\pi G = 1\,,$ and we will 
use the conventions and definitions introduced in~\cite{KSHM} unless stated
otherwise.

%%%%%%%%%%%%%%%%%%%%%%%%%%%%%%%%%%%%%%%%%%%%%%%%%%%%%%%%%%%%%%%%%%%%%%%%%%
%
%                SECTION TWO
%
%%%%%%%%%%%%%%%%%%%%%%%%%%%%%%%%%%%%%%%%%%%%%%%%%%%%%%%%%%%%%%%%%%%%%%%%%%

\section{On the consequences of the integrability conditions for the Killing
equations\label{alge}}

From now on we will assume the spacetime is endowed with a non-null KVF 
$\mb{\xi}$
\begin{equation}
N = \mbox{g}_{ab}\xi^a\xi^b \neq 0\,. \label{norm}
\end{equation}
From the definition of a Killing symmetry it follows
that $\mb{\xi}$ satisfies the Killing equations
\begin{equation}
\pounds_{\mbox{\tiny \boldmath $\xi$}}\mbox{g}_{ab} = 0~~ \Longleftrightarrow~~   
\xi_{a;b}+\xi_{b;a} = 0\,, \label{kill}
\end{equation}
where the semicolon denotes covariant differentiation.  We can know about
the Killing symmetries that a given spacetime admits by solving these
equations.  Of course, there are spacetimes that do not admit any
Killing symmetry, in which case one finds that the Killing equations
have no solutions.  We can investigate whether or not the Killing
equations admit solutions in a given spacetime by studying their 
integrability conditions, which can be expressed in the following 
compact form
\begin{equation}
\pounds_{\mbox{\tiny \boldmath $\xi$}}\Gamma^a{}_{bc} = 
\pounds_{\mbox{\tiny \boldmath $\xi$}}R^a{}_{bcd} =
\pounds_{\mbox{\tiny \boldmath $\xi$}}R^a{}_{bcd;e_1} = 
\pounds_{\mbox{\tiny \boldmath $\xi$}}R^a{}_{bcd;e_1e_2}
= \cdots = 0 \,, \label{uico}
\end{equation}
where $\pounds_{\mbox{\tiny \boldmath $\xi$}}$ denotes Lie differentiation 
along $\mb{\xi}$, $\Gamma^a{}_{bc}$ are the Christoffel symbols, and
$R^a{}_{bcd}$ the components of the Riemann curvature tensor.  As we can 
see, the integrability conditions involve, in general, derivatives of the
curvature tensor.

In recent works~\cite{FASO1,FASO2} we introduced an alternative way of
dealing with vacuum spacetimes admitting a Killing symmetry.  The main
object in this approach is the exterior derivative of the KVF
\[ \mb{F} = \mb{d\xi}~~~\Rightarrow~~~F_{ab} = \xi_{b;a}-\xi_{a;b}
= 2\xi_{b;a} \,,\]
where we have used the Killing equations~(\ref{kill}).  This quantity
is important for several reasons.  As it was firstly recognized
by Papapetrou~\cite{PAPA}, a KVF $\mb{\xi}$ can always be seen as the vector
potential generating an electromagnetic field, $F_{ab}$, satisfying 
Maxwell equations
\begin{equation}
F_{[ab;c]} = 0, \quad F^{ab}{}_{;b} = J^a \,, \label{maxe}
\end{equation}
where $J^a$ is the conserved current given by
\begin{equation}  
J^a = 2R^a{}_b\,\xi^b ~~ \Rightarrow ~~ J^a{}_{;a} = 0 \,, \label{sources}
\end{equation}
being $R_{ab} = R^c{}_{acb}$ the Ricci tensor.  Apart from
the Maxwell equations, $F_{ab}$ satisfies the following relation
\begin{eqnarray}
\pounds_{\mbox{\tiny \boldmath $\xi$}} F_{ab} = 0 \,, \label{inva}
\end{eqnarray}
which is a direct consequence of the Killing equations and it is
equivalent to
\begin{equation}
\xi^c F_{ab;c} = 0\,. \label{extra}
\end{equation}

In the approach proposed in~\cite{FASO1,FASO2} the components of the
Papapetrou field $F_{ab}$ are promoted to the level of variables of 
the formalism, like the components of the KVF.  Then, taking into 
account the Killing equations~(\ref{kill}), the equations for $\mb{\xi}$ 
are now
\begin{equation}
\xi_{b;a} = \textstyle{1\over2}F_{ab}~~~~~
\mbox{with}~~~~~F_{ab} = F_{[ab]} \,. \label{defpap}
\end{equation}
In the~\ref{appa} we have expressed them using the NP formalism.

Their integrability conditions are equivalent to the Ricci identities for
the KVF $\mb{\xi}$
\begin{equation} 
\xi_{a;dc}-\xi_{a;cd} = R_{abcd}\xi^b \,. \label{ricid}
\end{equation}
In the four-dimensional spacetimes of general relativity, and taking
into account only the antisymmetry of the Riemann tensor in the 
first and second pair of indices ($R_{abcd} = R_{[ab]cd} = R_{ab[cd]}$),
expression~(\ref{ricid}) contains 24 independent equations.  They can be
split into four differentiated groups: (i) First Bianchi identities:
$R_{[abc]d}\xi^d = 0\Rightarrow\xi_{[ab;c]} = 0$. They contain four 
independent equations which lead to the first group of Maxwell 
equations~(\ref{maxe}).  (ii) The trace: $\xi^a{}_{b;c}-\xi^a{}_{a;b} =
R_{bd}\xi^d$.  Four independent components equivalent to the
second set of Maxwell equations~(\ref{maxe}).  (iii) $R_{abcd}\xi^a
\xi^b = 0\Rightarrow\xi^a(\xi_{a;dc}-\xi_{a;cd}) = 0$.  This
expression contains 6 independent equations, which in combination
with those in the group (i), lead to the property~(\ref{inva},\ref{extra}) 
of $F_{ab}$.  (iv) The rest, 10 independent equations.  
Introducing~(\ref{defpap}) for the derivatives of $\mb{\xi}$ and the 
decomposition of the Riemann tensor into its irreducible parts under the 
Lorentz group: the Weyl tensor $C_{abcd}$, the curvature tensor part not 
determined locally by the energy-momentum distribution, the traceless Ricci 
tensor $S_{ab} = R_{ab}-\textstyle{1\over4}\mbox{g}_{ab}R$, and the scalar 
curvature $R = \mbox{g}^{ab}R_{ab}$, 
\[ R_{abcd} = C_{abcd}~+~\mbox{g}_{a[c}S_{d]b} -
\mbox{g}_{b[c}S_{d]a}~+~\textstyle{1\over12}(\mbox{g}_{ac}
\mbox{g}_{bd}-\mbox{g}_{ad}\mbox{g}_{bc})R \,,\]
the remaining 10 equations in~(\ref{ricid}) can be written in the following 
form
\begin{equation}
C_{abcd}\xi^d = - \textstyle{1\over2}F_{ab;c} + \xi_{[a}
R_{b]c}+\xi^dR_{d[a}\mbox{g}_{b]c}-\textstyle{1\over3}
\xi_{[a}\mbox{g}_{b]c}R  \,. \label{icond}
\end{equation}
The right-hand side is made out of a term constructed only from the
Papapetrou field and terms constructed from the Ricci tensor.
The last ones can be written in terms of the matter variables,
described by the energy-momentum tensor $T_{ab}\,,$ 
through the Einstein field equations
\begin{equation} 
G_{ab} = R_{ab}-\textstyle{1\over2}\mbox{g}_{ab}R = T_{ab} \,, \label{efe}
\end{equation}
where $G_{ab}$ denotes the Einstein tensor.  Then, we can think of 
equation~(\ref{icond}) as a relation that will provide expressions 
for some components of the Weyl tensor in terms of $F_{ab}$ and
$T_{ab}$.  As we have said before, equation~(\ref{icond}) contains 
10 independent equations, the same number of independent components 
of the Weyl tensor.  Hence, we can determine all the components of this
tensor from~(\ref{icond}).  However, this is true only in 
four-dimensional spacetimes.   For D-dimensional spacetimes we would
get an equation analogous to~(\ref{icond}), with the same left-hand
side and with the same terms on the right-hand side but with different 
numerical coefficients (see, e.g.,~\cite{HAEL} to obtain these 
coefficients).  The number of independent components of the Weyl
tensor in $D$ dimensions is (see, e.g.,~\cite{TOCHO})
\[ C(D) = \textstyle{1\over12}D(D+1)(D+2)(D-3)~~~~~(\mbox{for~}
D\geq 3) \,. \]
On the other hand, to find the number of independent components
in the D-dimensional version of equation~(\ref{icond}) we can write 
it as $C_{abcd}\xi^d = A_{abc}$.  Then, taking into account the 
information coming from (i)-(iii), the tensor $A_{abc}$ satisfy the 
following properties
\begin{equation}
A_{[abc]} = 0\,,~~~~~A_{[ab]c} = A_{abc}\,,~~~~~A_{abc}\xi^c = 0 
\,,~~~~~A_{ab}{}^a = 0\,. \label{prop}
\end{equation}
From these relations and the fact that we are considering KVFs with a 
non-vanishing norm $N$ [see Eq.~(\ref{norm})], the number of 
independent components contained in Eq.~(\ref{icond}) are
\[ I(D) = \textstyle{1\over6}D(D+1)(2D-5) \,. \]
Hence, the number of independent components of the Weyl tensor
that {\em cannot} be determined from the integrability
conditions~(\ref{icond}) is given by
\[ \Delta(D) = C(D) - I(D) = \textstyle{1\over12}D(D^2-1)(D-4) \,. \]
Therefore, in the case of General Relativity, $D = 4$, we can determine
completely the Weyl tensor [$\Delta(4) = 0$] in terms of the
Papapetrou field $F_{ab}$ and the energy-momentum content [see
Eq.~(\ref{icond})].  In higher dimensions\footnote{For $D\leq 3$
the Weyl tensor is identically zero.} this is not true, and we can
only determine a subset of the independent components of the Weyl
tensor.

As a conclusion we can say that from the integrability conditions~(\ref{icond}) 
and in $D = 4$ we can find an expression for the Weyl tensor in terms of the 
Papapetrou field and the energy-momentum content.    We can find the 
explicit expression of the Weyl tensor by different means.  Here we 
will use a procedure based on the fact that for a given non-null vector 
field we can associate with it what are called the electric and magnetic 
parts of the Weyl tensor.  We are interested in the electric and magnetic 
parts associated with the KVF $\mb{\xi}\,$:
\begin{eqnarray}
E_{ab}[\mb{\xi}] = C_{acbd}\xi^c\xi^d  \nonumber \\
H_{ab}[\mb{\xi}] = \ast C_{acbd}\xi^c\xi^d~~~~~(\ast C_{abcd} =
\textstyle{1\over2}\eta_{ab}{}^{ef}C_{cdef}) \,, \label{elma}
\end{eqnarray}
where $\ast$ denotes the dual operation and $\eta_{abcd}$ the components
of the completely antisymmetric volume four form.  The two 
tensors~(\ref{elma}) are symmetric, trace-free and orthogonal to the vector 
field they are associated with, the KVF in our case,
\[ \fl E_{(ab)}[\mb{\xi}] = E_{ab}[\mb{\xi}]\,,~H_{(ab)}[\mb{\xi}] =
H_{ab}[\mb{\xi}]\,,~~~E^a{}_a[\mb{\xi}] = H^a{}_a[\mb{\xi}] = 0
\,, ~~~E_{ab}[\mb{\xi}]\xi^b = H_{ab}[\mb{\xi}]\xi^b = 0\,. \]
Moreover, they contain all the information about the Weyl tensor.  Indeed,
each part has five independent components and the Weyl tensor has ten.
We can reconstruct the Weyl tensor from the electric and magnetic
parts and the KVF $\mb{\xi}$ through the following expression
\begin{equation}
\fl C_{abcd} = \frac{1}{N^2}\xi^e\xi^g\left\{ (\mbox{g}_{abef}
\mbox{g}_{cdgh}-\eta_{abef}\eta_{cdgh})E^{fh}[\mb{\xi}]-
(\mbox{g}_{abef}\eta_{cdgh}+\eta_{abef}\mbox{g}_{cdgh})
H^{fh}[\mb{\xi}]\right\} \,, \label{weyl}
\end{equation}
where $\mbox{g}_{abcd} = \mbox{g}_{ac}\mbox{g}_{bd}-\mbox{g}_{ad}
\mbox{g}_{bc}\,.$

From the integrability conditions~(\ref{icond}) we can compute
the electric and magnetic parts associated with the KVF $\mb{\xi}$,
the result is
\begin{eqnarray}
E_{ab}[\mb{\xi}] = -\textstyle{1\over2}\xi^e\mbox{g}_{ea}{}^{cd}A_{cdb}
\,, \label{elec} \\
H_{ab}[\mb{\xi}] = -\textstyle{1\over2}\xi^e\eta_{ea}{}^{cd}A_{cdb}
\,, \label{magn}
\end{eqnarray}
where the tensor $A_{abc}$ was introduced above and has the properties
shown in~(\ref{prop}).  In terms of $F_{ab}$ and $T_{ab}$ it has the
following expression
\begin{equation} 
A_{abc} = - \textstyle{1\over2}F_{ab;c} + \xi_{[a}
T_{b]c}+\xi^dT_{d[a}\mbox{g}_{b]c}-\textstyle{2\over3}
\xi_{[a}\mbox{g}_{b]c}T  \,, \label{defa}
\end{equation} 
where $T = \mbox{g}^{ab}T_{ab}\,.$
Then, introducing this expression in
equations~(\ref{elec},\ref{magn}) and finally into the Weyl tensor
[Eq.~(\ref{weyl})] we get
\begin{equation}
 C_{ab}{}^{cd} = \frac{2}{N}\left\{ A_{ab}{}^{[c}\xi^{d]}
+A^{cd}{}_{[a}\xi_{b]}+2\xi^eA_{e}{}^{[c}{}_{[a}\delta^{d]}{}_{b]}
\right\} \,. \label{cccc}
\end{equation}
And this is how the integrability conditions for the components of the
KVF supply an expression for the Weyl tensor in terms of the KVF,
the Papapetrou field and the energy-momentum tensor.

We can insert~(\ref{cccc}) into the contracted second Bianchi 
identities (which in $D = 4$ contain the same information as 
the non-contracted ones)
\[ C_{abcd}{}^{;d} =  R_{c[a;b]}- \textstyle{1\over6} 
\mbox{g}_{c[a}R_{,b]} \,.\] 
The result, using Einstein's equations~(\ref{efe}), is
\begin{equation} 
F_{ab;c}{}^c + C_{abcd} F^{cd}+ 2 J_{[a;b]}+ 
\textstyle{1\over3} T F_{ab} = 0\,. \label{deri} 
\end{equation}
One can reach this result starting from Maxwell equations~(\ref{maxe}),
taking the covariant derivative of $F_{[ab;c]} = 0$ and using the
Ricci identities.  This means that~(\ref{deri}) is an identity 
as long as $F_{ab}$ satisfies Maxwell equations.

%%%%%%%%%%%%%%%%%%%%%%%%%%%%%%%%%%%%%%%%%%%%%%%%%%%%%%%%%%%%%%%%%%%%%%%%%%
%
%                SECTION THREE
%
%%%%%%%%%%%%%%%%%%%%%%%%%%%%%%%%%%%%%%%%%%%%%%%%%%%%%%%%%%%%%%%%%%%%%%%%%%

\section{On the Integrability conditions for the Papapetrou field
and other issues\label{papi}}

In the previous section we have studied the integrability conditions
for the components of the KVF.  The next step is to study the 
equations for the Papapetrou field $F_{ab}$, i.e., Maxwell 
equations~(\ref{maxe}).  Their integrability conditions have been
studied in great detail in~\cite{CFF}.  Here, we will apply their 
results to the case of a Papapetrou field.  To that end, it is very 
convenient to consider the following two points.  First,
instead of using $F_{ab}$, we will work with the self-dual Papapetrou
field
\[ \tilde{F}_{ab} = F_{ab} + i\ast F_{ab} ~~~~~ (\ast F_{ab} =
\textstyle{1\over2}\eta_{abcd} F^{cd})\,. \] 
This will lead to more compact expression in our study.  In particular,
the Maxwell equations look simply as
\begin{equation}
\tilde{F}^{ab}{}_{;b} = J^a \,, \label{ME} 
\end{equation}
where $J^a$ are the conserved sources, given in~(\ref{sources}).
The second point is to take into account the algebraic structure 
of the Papapetrou field, which in~\cite{FASO1,FASO2} has been shown to be 
fundamental.  The idea is to distinguish between the two possible 
algebraic types of $F_{ab}$: (i) The {\em regular} type, characterized by
$\tilde{F}^{ab}\tilde{F}_{ab}\neq 0\,.$  In this case we can pick a 
Newman-Penrose basis \npb so that $\mb{\tilde{F}}$ takes the following 
{\em canonical} form
\begin{equation}
\fl \tilde{F}_{ab} = \Upsilon  W_{ab} \,, ~~ \mbox{where} ~~
\Upsilon = -(\al\plusi\bb) ~~\mbox{and}~~
W_{ab} = -2k_{[a}\ell_{b]}+2m_{[a}\bar{m}_{b]} \,,  \label{regc}
\end{equation}
and where \al~and \bb~are the real eigenvalues of $F_{ab}$, and $\mb{k}$ and 
$\mb{\ell}$ are its null eigenvectors (principal null directions).  
(ii) The {\em singular} type, characterized 
by $\tilde{F}^{ab}\tilde{F}_{ab} = 0$.  Now, we can choose the Newman-Penrose
basis so that $\mb{\tilde{F}}$ can be cast in the form
\begin{equation}
\fl \tilde{F}_{ab} = 2\theta V_{ab}\,, ~~\mbox{where}~~ 
V_{ab} = 2k_{[a}m_{b]}\,,\label{sinc}
\end{equation}
where $\theta$ is a complex scalar and $\mb{k}$ gives the only principal 
null direction.  

The NP basis in~(\ref{regc},\ref{sinc}) are not completely fixed. 
The transformations that keep $\tilde{F}_{ab}$ in the canonical
form were discussed in~\cite{FASO2}.  We only mention the 
fact that whereas in the regular case the eigenvalues 
\al~and \bb~are invariant under these transformation, in the singular
case the scalar $\theta$ is not, actually we can scale it in an arbitrary
way.  

We can now study the integrability conditions of the Maxwell
equations for the Papapetrou field~(\ref{ME}).  We are 
going to consider each algebraic case separately.

%%%%%%%%%%%%%%%%%%    SUBSECTION THREE.ONE    %%%%%%%%%%%%%%%%%%%%%%%

\subsection{Regular case}
Taking into account the following property of $W_{ab}$
\begin{equation} 
W_a{}^b W_{bc} = \mbox{g}_{ac} \,, \label{prow}
\end{equation}
one can show that the Maxwell equations~(\ref{ME}) are equivalent
to the following equations
\begin{equation}
\Upsilon_{,a} = \Upsilon h_a + q_a \,, \label{phic}
\end{equation}
where
\begin{equation}
h_a =  W_{ab;c} W^{bc} \,, ~~  
q_a = W_{ab}J^b = 2 W_a{}^b R_{bc} \xi^c \,. \label{h}
\end{equation}
Remarkably, Maxwell equations provide an expression for all the 
derivatives of the eigenvalues \al~ and \bb~(\ref{phic}) [their 
expressions in the NP formalism are given in the \ref{appa}, 
Eqs.~(\ref{maxr1}-\ref{maxr4})].  In this case the integrability 
conditions for the Maxwell equations reduce to the integrability 
conditions for the complex scalar $\Upsilon\,.$  They will impose conditions 
on $h_a$ and $q_a$.  Following~\cite{CFF}, we divide the problem into 
two cases: (a) $\mb{dh} = 0 \Rightarrow h_{[a,b]} = 0$. 
(b) $h_{[a,b]}\neq 0$. Case (a) represents a particular situation in 
which a condition on  the 2-form $W_{ab}$ has been imposed.  
The integrability conditions for the complex scalar $\Upsilon$ are simply
\begin{eqnarray}
\mb{dq}+\mb{q}\wedge\mb{h} = 0~~\Longrightarrow~~ 
{q}_{[a,b]} + {q}_{[a} h_{b]} = 0\,. \label{intph}
\end{eqnarray}
This condition is automatically satisfied when $q_a = 0$, 
or equivalently when $R_{ab}\xi^b = 0$ (the KVF is an eigenvector 
of the Ricci tensor with zero eigenvalue), which includes the 
vacuum case.  The integrability conditions~(\ref{intph}) 
are first-order equations for $\mb{q}$ and their integrability
conditions, $\mb{d^2q} = 0\,,$ are identically satisfied by 
virtue of $\mb{dh} = 0\,.$  The expressions of this last 
set of equations in the NP formalism is given in~\ref{appa}, 
Eqs.~(\ref{intmax1}-\ref{intmax5}).
Case (b) constitutes the generic situation.  The integrability
conditions for $\Upsilon$ can be written as follows (see~\cite{CFF}
for details)
\begin{eqnarray}
\Omega_{ab} h_{[c,d];e} - h_{[a,b]}[ \Omega_{cd;e} - h_c
\Omega_{de} + q_c h_{[d,e]}] = 0\,, \label{genint}
\end{eqnarray}
where
\begin{eqnarray}
\Omega_{ab} = q_{[a,b]} + q_{[a} h_{b]}\,. \nonumber
\end{eqnarray}

Another important issue that we must consider is the analysis of the 
consequences of the relation~(\ref{extra}).  To that end it is useful
to use the complex 2-forms $\{\mb{U},\mb{V},\mb{W}\}\,,$ where the 
2-form $\mb{U}$ is given by
\[ U_{ab} = -2l_{[a}\bar{m}_{b]}\,, \]
and $\mb{W}$ and $\mb{V}$ have been introduced in~(\ref{regc})
and~(\ref{sinc}) respectively.  Using~(\ref{phic}), 
the relation~(\ref{extra}) leads to the following three equations
\begin{eqnarray}
U^{ab}\xi^cW_{ab;c} = 0 \,, \label{extr1}                     \\
V^{ab}\xi^cW_{ab;c} = 0 \,, \label{extr2}                     \\
\xi^a(\Upsilon h_a+q_a) = 0 \Longrightarrow \xi^a\Upsilon_{,a} = 0 \,. 
\label{extr3}
\end{eqnarray}
These relations mix algebraically the components of the KVF with
components of the connection and the Ricci tensor.  We have
given their explicit form in the NP formalism in~\ref{appa}
[Eqs.~(\ref{invr1}-\ref{invr3})].

Finally, from~(\ref{elec},\ref{magn}) and~(\ref{magn}) one can construct
the invariants of the Weyl tensors in terms of the Papapetrou field
and the energy-momentum tensor.   In the case of vacuum spacetimes
and a timelike KVFs, one can also construct the Bel-Robinson 
{\em superenergy} (see~\cite{BEL,MANSEN}) associated with the observers 
following the trajectories of the KVF.  These quantities have been
used in numerical simulations to track, e.g., the propagation of 
gravitational waves.  In the regular case and vacuum, 
by virtue of~(\ref{phic}), the Bel-Robinson superenergy has the form:   
$(\al^2+\bb^2)^2 \mbox{(Connection terms)}^2$.   For the nonvacuum
case, the Bel-Robinson tensor is not longer conserved, but there
are generalizations.  In the case of spacetimes with vanishing
scalar curvature, which include Einstein-Maxwell spacetimes, we have 
a generalization~\cite{MANME} which also has a positive timelike 
component (superenergy), where apart from terms of the form showed 
before we have terms constructed from the energy-momentum tensor
and mixed terms of the form: 
$(\al^2+\bb^2)\mbox{(Connection terms)}\mbox{(Components of $T_{ab}$)}$.

%%%%%%%%%%%%%%%%%%    SUBSECTION THREE.TWO    %%%%%%%%%%%%%%%%%%%%%%%%

\subsection{Singular case}

The situation in the singular case is completely different.
The main reason is that $V_{ab}$ is a singular 2-form
which obviously does not satisfy~(\ref{prow}).  The main consequence of
this fact is that the Maxwell equations~(\ref{ME}),
which now look as follows 
\begin{equation}
V_a{}^b\theta_{,b} = -\theta V_a{}^b{}_{;b} + 
\textstyle{1\over2}J_a\,. \label{varphic}
\end{equation}
does not provide expressions for all the derivatives of 
the complex scalar $\theta$ [see~\ref{appa2}].  Therefore,
we can only talk about the compatibility conditions for
the expressions of the known derivatives of $\theta\,.$ 
Actually, as we have previously said, contrary
to what happens with $\Upsilon$, $\theta$ is not an invariant quantity,
and we can even scale it arbitrarily.   Then, the question of the
compatibility conditions is not a key issue.  

Finally, the consequences of the relation~(\ref{extra})
lead only to two (instead of three as in the regular case)
complex equations
\begin{eqnarray}
\xi^c\theta_{,c} +\textstyle{1\over2}\theta U^{ab}\xi^cV_{ab;c} = 0 
\,, \label{exts1} \\
W^{ab}\xi^cV_{ab;c} = 0 \,. \label{exts2}
\end{eqnarray}
The second one is the same as the second one in the regular case.
We have given their explicit form in the NP formalism in~\ref{appa}
[Eqs.~(\ref{invs1},\ref{invs2})].

%%%%%%%%%%%%%%%%%%%%%%%%%%%%%%%%%%%%%%%%%%%%%%%%%%%%%%%%%%%%%%%%%%%%%%%%%%
%
%                SECTION FOUR
%
%%%%%%%%%%%%%%%%%%%%%%%%%%%%%%%%%%%%%%%%%%%%%%%%%%%%%%%%%%%%%%%%%%%%%%%%%%

\section{General framework\label{genf}}

We have seen how the introduction of the Papapetrou field allows us to gain 
new insights into the study of the consequences of an isometry in spacetime 
local structure.  This was already considered in~\cite{FASO1,FASO2} for 
the case of vacuum spacetimes.  There, the Papapetrou field was used to set 
up a classification of these spacetimes and to study the possible relations 
between the existence and structure of a Killing symmetry and the algebraic
classification of spacetimes, the Petrov classification~\cite{PETROV,KSHM}.
Moreover, a new formalism was set up, which provides control both on the
structure of the Killing symmetry and on the algebraic structure of the 
spacetime.  The way this formalism works was shown in~\cite{FASO2} and a 
number of results were derived (see~\cite{STEELE1} for the correction of one 
of the results).   In what follows we will set up a framework which takes 
into account all the information we have presented in the previous sections, 
and which extends the formalism developed in~\cite{FASO2} for vacuum 
spacetimes to any spacetime, without restrictions on the matter content.

The main idea is to extend the Newman-Penrose (NP) formalism by adding 
quantities, and their corresponding equations, that describe the Killing 
symmetry and its associated Papapetrou field.   Briefly speaking, the 
Newman-Penrose formalism~\cite{NEPE} is a tetrad formalism where all 
the tetrad vectors are null vectors.  It has reveal itself as a very
powerful tool in many different tasks (e.g., exact solutions, study
of gravitational radiation, etc.), an in 
particular, using it we get control on the algebraic structure of the 
spacetime.   The variables and equations in this formalism are 
(see~\cite{KSHM,PAP1,PAP2} for definitions and other details):
(i) The components of a NP basis $(z_a{}^b) = (k^b,\ell^b,m^b,\bar{m}^b)$
in a coordinate system $\{x^a\}$.
The equations for them come from the definition of the associated
connection.  We obtain them by applying the commutators of the NP basis 
vectors (equations~(7.55)-(7.58) in~\cite{KSHM}) to the coordinate
system $\{x^a\}$. (ii)  The components of the connection, known as
spin coefficients.  They are 12 complex scalars named as: ($\kappa$,
$\sigma$, $\rho$, $\epsilon$, $\nu$, $\lambda$, $\mu$, $\gamma$,
$\tau$, $\pi$, $\alpha$, $\beta$).  The equations they obey come from
the integrability conditions for the variables in (i), the so-called NP 
equations (equations~(7.28)-(7.45) in~\cite{KSHM}), which are just
the expressions of the components, in the NP basis, of the 
Riemann tensor in terms of the complex connection. 
(iii) Riemann curvature variables.  These are divided into
Weyl tensor components, described in the NP formalism by five complex scalars
$\Psi_A$ ($A = 0,\ldots,4$); components of the traceless Ricci tensor,
given by complex scalars $\Phi_{XY}$ ($X,Y = 0,1,2$) satisfying 
$\Phi_{XY} = \bar{\Phi}_{YX}$; and the scalar curvature, $\Lambda$.  
The equations for these quantities are the second Bianchi identities 
(equations~(7.61)-(7.71) in~\cite{KSHM,NOTE}), which are at the 
same time integrability conditions for the NP equations.  
With these equations we get a closed system of equations for 
the whole set of variables.   Finally, when we get a specific 
description of the energy-momentum content through a set of matter fields, 
we can express the quantities $\Phi_{XY}$ and $\Lambda$ in terms of 
these fields and to add the equations for the matter fields.  
A good example is the case of electromagnetic fields (see, e.g.,~\cite{KSHM}).

Our formalism is an extension of the NP formalism based on the following
two ideas:  (A) To adapt the NP formalism as much as possible to the 
existence of a Killing symmetry.  (B) To extend the NP formalism by 
adding new variables and equations characterizing the Killing symmetry.

The best way of implementing the point (A) is through the choice of the
NP basis \npb.  We will choose the NP basis to be adapted to the 
algebraic structure of the Killing symmetry.  That is, in the case
regular case we will choose a NP basis in which the Papapetrou field
$\mb{F}$ takes the canonical form~(\ref{regc}), and in the singular
case we choose a NP basis in which $\mb{F}$ has the canonical 
form~(\ref{sinc}).   Then, in our framework all the quantities and 
equations of the NP formalism will be written in such an adapted NP 
basis.   With regard to point (B), we extend the NP formalism by adding,
to the sets described above [(i)-(iii)], the following variables and 
equations that describe the Killing symmetry:
(iv) The components of the KVF in the chosen adapted NP basis.
They are introduced through the following expressions
\begin{equation} 
\xi_k = k^a\xi_a\,, \hspace{5mm}
\xi_l = \ell^a\xi_a\,, \hspace{5mm}
\xi_m = m^a\xi_a\,, \hspace{5mm}
\xi_{\bar{m}} = \bar{m}^a\xi_a\,, \label{ckvf}
\end{equation}
where $\xi_k$ and $\xi_l$ are real scalars, and $\xi_m$ and $\xi_{\bar{m}}$ 
are complex and related by $\xi_{\bar{m}} = \bar{\xi}_m\,.$  
Then,  
\begin{eqnarray}
\mb{\xi} = -\xi_l\mb{k}-\xi_k\mb{\ell}+\bar{\xi}_m\mb{m}+
\xi_m \mb{\bar{m}}\,.  \label{kvfnp}
\end{eqnarray}  
The equations for the variables $(\xi_k,\xi_l,\xi_m)$ are obtained from
the definition of the Papapetrou field~(\ref{defpap}).  Their form once 
projected onto an adapted NP basis is shown in 
equations~(\ref{ckvf1}-\ref{ckvf10}) of~\ref{appa1}.  (v) The components 
of the Papapetrou field in the adapted NP basis.  In the regular case 
these components are completely described by the complex quantity 
$\Upsilon = -(\al\plusi\bb)$, and in the singular case
by the complex scalar $\theta\,.$  The equations for these quantities come
from the projection of Maxwell equations~(\ref{maxe}) onto the adapted
NP basis.   We have given them in~\ref{appa2}.  The equations for
the regular case are~(\ref{maxr1}-\ref{maxr4}), and the equations
for the singular case are~(\ref{maxs1}-\ref{maxs4}).

Another set of equations that we have to take into account to
complete the formalism are those coming from Eqs.~(\ref{inva},\ref{extra}).
As we have seen before, in the regular case they lead to 3 complex 
algebraic equations [Eqs.~(\ref{extr1}-\ref{extr3})].  When we
express the equations in the NP formalism, using a basis adapted
to the Papapetrou field, they become equations involving the components 
of the Killing, the spin coefficients and components of the Ricci
tensor [see Eqs.~(\ref{invr1}-\ref{invr3}) in~\ref{appa}].  
In the singular case we only get two complex equations
[Eqs.~(\ref{exts1},\ref{exts2})], and one of them, Eq.~(\ref{exts1}),
coincides with one of the regular case, namely Eq.~(\ref{extr2}).
Their expressions in the NP formalism are similar to those in
the regular case, with the only difference that now they involve
derivatives of $\theta$ [see Eqs.~(\ref{invs1},\ref{invs2}) in~\ref{appa}].

So far we have presented the formalism, now let us see what is the best
way of using it.  Following the discussion in section~\ref{alge}, 
the integrability conditions for the components of the KVF are a good
starting point since they lead to an expression for the Weyl tensor 
[see Eqs.~(\ref{defa},\ref{cccc})].  That means that we can express
the complex scalars $\Psi_A$ ($A = 0,\ldots,4$) in terms of the rest
of quantities of the formalism.   In~\ref{appa3} we give these expressions 
for the regular [Eqs.~(\ref{rpsi0}-\ref{rpsi4})]
and singular [Eqs.~(\ref{spsi0}-\ref{spsi4})] cases.

Remarkably, the expressions for the regular case are algebraic in 
the spin coefficients, the eigenvalues of the Papapetrou field
$\al+i\bb$, the KVF components, and the components of the
Ricci tensor.  The reason for that comes from the Maxwell
equations, which are equivalent to the equations~(\ref{phic}).
Then, the quantity $F_{ab;c}$ can be written in terms of
$F_{ab}$, the connection and the Ricci tensor, hence the
algebraic expressions for $\Psi_A$.  In the singular case
some derivatives of the complex scalar $\theta$ appear in
the Weyl tensor but, as we have mentioned before, we can
arbitrarily scale this quantity by using the remaining freedom 
in the choice of the NP basis.

The main consequence of having an expression for the Weyl tensor
is that we do not need the equations for it.  Then, from the
second Bianchi identities we only need to consider the 
subset corresponding to the energy-momentum conservation
equations
\begin{equation} 
G^{ab}{}_{;b} = 0~~\Rightarrow~~T^{ab}{}_{;b} = 0\,. \label{csbi}
\end{equation}
Then, the situation after introducing the information coming
from the integrability conditions for the components of the 
KVF is a simplified framework, where we can forget about the 
components of the Weyl tensor and their equations.   

Apart from the equations we have listed above, we must take into
account the integrability/compatibility conditions for the
components of the Papapetrou field $F_{ab}\,,$ 
and the consequences of the fact that $F_{ab}$ is invariant under 
the 1-parameter family of diffeomorphisms generated by the KVF 
$\mb{\xi}$ [Eqs.~(\ref{inva},\ref{extra})].  The first group of
equations have been considered in section~\ref{papi}.  In
the particular case when the Papapetrou field is regular
and $\mb{dh} = 0$, we have seen that they lead to five independent
equations only for the connection, the spin coefficients, 
which complement the Newman-Penrose equations.    
We have given them in the \ref{appa} 
[Eqs.~(\ref{intmax1}-\ref{intmax5})].  In the general regular
case the integrability conditions are~(\ref{genint}).
When the Papapetrou field is singular, we can only talk about
compatibility conditions.

Finally, in the case that we know the particular description
for the energy-momentum content, that is, the matter fields and
the field equations they satisfy, we can adopt the following
point of view.  First, we can add the matter fields as new 
variables of the formalism, and the field equations they obey as 
equations of the formalism.  Then, we will get an expression,
via the energy-momentum tensor, for the components of the Ricci 
tensor, $\Phi_{XY}\,,\Lambda$, in terms of the matter fields and, 
probably, of their derivatives.  The advantage of this is that now 
we can ignore the quantities $\Phi_{XY}$ and the equations 
that they satisfy, namely the contracted second Bianchi 
identities~(\ref{csbi}).  The last point would be to consider
the integrability/compatibility conditions for the equations
of the matter fields, in a similar way as we have done with 
the equations for the Papapetrou field in the previous section.
In the next section, we illustrate this alternative procedure
with the case of spacetimes with a gravitating electromagnetic field.

%%%%%%%%%%%%%%%%%%%%%%%%%%%%%%%%%%%%%%%%%%%%%%%%%%%%%%%%%%%%%%%%%%%%%%%%%%
%
%                SECTION FIVE
%
%%%%%%%%%%%%%%%%%%%%%%%%%%%%%%%%%%%%%%%%%%%%%%%%%%%%%%%%%%%%%%%%%%%%%%%%%%

\section{Particular case: gravitating electromagnetic field\label{eima}}

To illustrate the framework we have described above we are going to 
consider the particular case of spacetimes whose energy-momentum
content corresponds to an electromagnetic field, i.e., we have
a gravitating electromagnetic field.  We will call it $H_{ab}$,
to distinguish it from the Papapetrou field $F_{ab}$.  The
self-dual field $\tilde{H}_{ab}$ will satisfy Maxwell equations
\begin{eqnarray}
\tilde{H}^{ab}{}_{;b} & = & S^a  \,,\label{MEH} 
\end{eqnarray}
where $S^a$ are the electromagnetic sources.  In general, 
$H_{ab}$ will not be aligned with the Papapetrou field, so
in the NP basis adapted to the Papapetrou field it will
take the following generic form
\begin{eqnarray}
\tilde{H}_{ab} & = & \Phi_0 U_{ab} + \Phi_1 W_{ab} + \Phi_2 V_{ab}
\,, \label{hem}
\end{eqnarray}
where $\Phi_{X}$ ($X = 0,1,2$) are the components of $\tilde{H}_{ab}$
in such a basis.  The energy-momentum tensor of the spacetime
will be given by
\begin{equation}
T_{ab} = \textstyle{1\over2} \tilde{H}_a{}^c \bar{\tilde{H}}_{cb}
~~\Rightarrow~~\mbox{g}^{ab}T_{ab} = 0\,. \label{TH}
\end{equation}
Therefore, the components of the Ricci tensor in terms of the
components of $H_{ab}$ are (see, e.g.,~\cite{KSHM})
\begin{equation}
\Phi_{XY} = \Phi_X \bar{\Phi}_Y\,, ~~~~~\Lambda = 0 \,. \label{phip}
\end{equation}
The NP equations for $\Phi_X$ can be found, e.g.,
in~\cite{KSHM} [Eqs.~(7.46)-(7.49)].  Following the framework
described above, we have to include the quantities 
$\Phi_X$ and their equations, but then we can forget about
$\Phi_{XY}$, which are given by~(\ref{phip}), and their
equations, the contracted second Bianchi 
identities~(\ref{csbi}).  So we have to concentrate on
the Maxwell equations for the gravitating electromagnetic
field.  If we insert~(\ref{hem}) into Maxwell equations~(\ref{MEH}) 
we obtain
\begin{eqnarray}
{\Phi_1}_{,a} = \Phi_1 h_a + \hat{q}_a \,, \label{Phi1c} \\
\hat{q}_a = W_{ab} S^b + W_{ac} (\Phi_0 U^{bc} + \Phi_2 V^{bc})_{;b} \,,
\nonumber
\end{eqnarray}
where $h_a$ is given in~(\ref{h}).  Again we can follow~\cite{CFF}
to study the integrability conditions for (\ref{Phi1c}), and
the classification used for the case of the Papapetrou field
depending on whether $h_{[a,b]}$ vanishes or not.  Here,
it is important to remark that this 1-form $h_a$ is the same
for both the Papapetrou field $F_{ab}$ and the gravitating
electromagnetic field $H_{ab}$.  This means we can analyze
simultaneously the integrability conditions for $F_{ab}$ 
(in the regular case) and $H_{ab}$.  Then, when $\mb{dh} = 0$ 
the integrability conditions are given by (\ref{intph}) and 
%\begin{equation}
\[ \hat{q}_{[a,b]} + \hat{q}_{[a} h_{b]} = 0\,. \]
%\end{equation}
The condition $\mb{dh} = 0$ takes place in the case of 
Einstein-Maxwell spacetimes ($S^a = 0$) where $F_{ab}$ and $H_{ab}$
are regular and aligned (see the particular case below in
the subsection~\ref{subsec51}).

In the generic case, when $h_{[a,b]}\neq 0$, the integrability 
conditions are given by~(\ref{genint}) and 
\begin{eqnarray}
\hat{\Omega}_{ab} h_{[c,d];e} - h_{[a,b]}[
\hat{\Omega}_{cd;e} - h_c
\hat{\Omega}_{de} + \hat{q}_c h_{[d,e]}] = 0\,, \nonumber 
\end{eqnarray}
where
\begin{eqnarray}
\hat{\Omega}_{ab} = \hat{q}_{[a,b]} +
\hat{q}_{[a} h_{b]} \,. \nonumber  %\label{Omdef}
\end{eqnarray}

%%%%%%%%%%%%%%%%%%    SUBSECTION FIVE.ONE    %%%%%%%%%%%%%%%%%%%%%%%%%

\subsection{A particular case: Einstein-Maxwell spacetimes 
with alignment between the Papapetrou and the electromagnetic field
\label{subsec51}}

Let us now consider a very particular case, in which the gravitating
electromagnetic field is source-free ($S^a = 0$) and it is 
aligned with the Papapetrou field in the following sense
\begin{equation} 
\tilde{H}_{ab} = \zeta\, \tilde{F}_{ab} \,, \label{alignment}
\end{equation}
where $\zeta$ is an arbitrary complex scalar. 
In the case of a regular Papapetrou field, $H_{ab}$ will have
two principal directions which coincide with those of $F_{ab}$.   
Moreover, we will have 
$\Phi_0 = \Phi_2 = 0$ in~(\ref{hem}), which means that now we 
have $\hat{q}_a = 0$.  Therefore, from~(\ref{phip}) it follows that
\[ \Phi_{00} = \Phi_{01} = \Phi_{02} = \Phi_{12} = \Phi_{22} = 0 = \Lambda
\,. \]
Or in other words, the only non-zero component of the Ricci
tensor is $\Phi_{11} = \Phi_1\bar{\Phi}_1$.  Comparing the equations
for the Papapetrou and the electromagnetic fields we get an
equation for the proportionality factor $\zeta$
\[ \mb{d}\mb{\log \zeta} = -\Upsilon^{-1}\mb{q} \,. \]
The integrability conditions for $\zeta$ are the same as those for 
$\Upsilon$, i.e. Eqs.~(\ref{intph}), corresponding to the case $\mb{dh} = 0$.
The interesting characteristic of this particular case is that if 
we look at the expressions for the Weyl tensor components, 
Eqs.~(\ref{rpsi0}-\ref{rpsi4}),  we see that they are formally the 
same as those for the vacuum case~\cite{FASO2}.  Also the NP equations 
coming from~(\ref{inva}), Eqs.(\ref{invr1}-\ref{invr3}), have the same 
form as in the vacuum case.  Finally, the particular case where
the proportional factor is constant, $\mb{d\zeta} = 0$, is empty since 
it implies either $\Upsilon = 0$ or $\zeta = 0$.

%%  EXAMPLE OF THE KERR-NEWMAN METRIC %%

An interesting example in which we find this kind of alignment is the 
case of the Kerr-Newman metric~\cite{KENE,KSHM}, which in
Boyer-Linquist coordinates reads
\begin{eqnarray}
\fl ds^2=-\frac{\Delta}{\varrho^2}(dt-a \sin^2\theta d\varphi)^2 +
\frac{\sin^2\theta}{\varrho^2}[(r^2 + a^2)d\varphi -a dt]^2 +
\frac{\varrho^2}{\Delta}dr^2 + \varrho^2 d\theta^2 \,, \nonumber \label{kn}
\end{eqnarray}
where $\Delta= r^2 - 2 M r+ a^2 + Q^2$ and $\varrho^2= r^2+ a^2
\cos^2 \theta$, being $M$, $a$ and $Q$ constants denoting respectively the 
mass, angular momentum, and charge of the Kerr-Newman black hole. 
It is easy to see that the Papapetrou field
associated with the timelike KVF $\mb{\xi_t}=\mb{\partial/\partial t}$ 
has the following form
\begin{eqnarray}
\mb{F} & = & \frac{2}{\varrho^4}[M(r^2 -a^2 \cos^2\theta)-r Q^2](\mb{dt} -
a \sin^2\theta \mb{d\varphi})\wedge \mb{dr}
\nonumber\\
&& + \frac{a(2Mr-Q^2)\sin(2\theta)}{\varrho^4}\mb{d\theta}\wedge(a \mb{dt} -
(r^2+a^2) \mb{d\varphi})\,. \label{fkn}
\end{eqnarray}
The Kerr-Newman metric is a solution of the Einstein-Maxwell
system of field equations, and the electromagnetic field $H_{ab}$ generating
the energy-momentum distribution is given by
\begin{eqnarray}
\mb{H}= -\sqrt{2}\,\frac{Q}{\varrho^4}(r^2-a^2 \cos^2\theta)(\mb{dt} -
a \sin^2\theta \mb{d\varphi})\wedge \mb{dr}\nonumber\\
- \sqrt{2}\,\frac{rQ}{\varrho^4} a \sin(2\theta)
\mb{d\theta}\wedge(a \mb{dt} - (r^2+a^2)\mb{d\varphi})\,, \label{hkn}
\end{eqnarray}
which can be generated by the following vector potential
\begin{eqnarray}
\mb{A} = -\sqrt{2}\,\frac{rQ}{\varrho^2}(\mb{dt}-a\sin^2\theta\mb{d\varphi})
= - \frac{\sqrt{2}\,rQ}{2Mr-Q^2}(\mb{\xi_t}+\mb{dt}) \,. \nonumber
\end{eqnarray}
From the expressions for $\mb{F}$ [Eq.~(\ref{fkn})] and $\mb{H}$
[Eq.~(\ref{hkn})] we can construct the respective self-dual 2-forms,
and from~(\ref{alignment}) we can find $\zeta$.  The result is
\begin{eqnarray}
\zeta=-\frac{Q}{\sqrt{2}}\frac{r+i a\cos\theta}{M(r+ia\cos\theta)-Q^2}
\,. \nonumber  
\end{eqnarray}
Since $\mb{F}$ and $\mb{H}$ are aligned in the sense of~(\ref{alignment})
they have the same principal null directions.  These directions,
which also coincide with the two multiple directions of the
spacetime (of the Weyl tensor), are given by the following vector
fields
\begin{eqnarray}
\fl \mb{K}= -\frac{\Delta}{\varrho^2}\mb{dt}+ \mb{dr} +\frac{a
\sin^2\theta \Delta}{\varrho^2}\mb{d\varphi}\,,\ \ \ \ \
\mb{L}=-\frac{\Delta}{\varrho^2}\mb{dt}- \mb{dr} +\frac{a
\sin^2\theta \Delta}{\varrho^2}\mb{d\varphi}\,. \nonumber
\end{eqnarray}
Note that these vector fields are not normalized, i.e. $K^aL_a\neq -1$.
From them we can compute the complex scalar $\Upsilon$ containing
the eigenvalues \al\, and \bb\, [see Eq.~(\ref{regc})]:
\begin{eqnarray}
\Upsilon= -2\frac{(r+i a \cos\theta)}{\varrho^4}\left[M(r+i a
\cos\theta)-Q^2\right] \,. \nonumber
\end{eqnarray}
Using all this information we can finally get $\mb{q}$ and $\mb{h}$:
\begin{eqnarray}
\mb{q}= -2\frac{Q^2}{\varrho^4} \mb{d(r + i a \cos \theta)}\,,
\hspace{5mm} \mb{h} = -2\mb{d\ln(r-ia\cos\theta)}\,.
\nonumber    
\end{eqnarray}
As expected, we have $\mb{dh}=0$.

\vspace{3mm}

%%  END OF THE KERR-NEWMAN EXAMPLE  %% 

In the case of a singular Papapetrou field, $H_{ab}$ will be also
singular and with the same principal direction.  In this case
we have $\Phi_0 = \Phi_1 = 0$ in~(\ref{hem}), hence we find that
\[ \Phi_{00} = \Phi_{01} =\Phi_{02} = \Phi_{11} = \Phi_{12} = 0 = \Lambda
\,. \]
Then, the only non-zero component of the Ricci tensor is
$\Phi_{22} = \Phi_2\bar{\Phi}_2$.  From the expressions for 
the $\Psi_A$ ($A = 0,...,4$) given in the~\ref{appa} we have
\[ \Psi_0 = \Psi_1 = \Psi_2 = 0 \,, \]
which means that the spacetime is not only algebraically special,
as the Mariot-Robinson theorem~\cite{MARI,ROBI} tells us, but also that it 
must be either of the Petrov type III or N.  In other words, the only
principal null direction, $\mb{k}$, must have at least multiplicity
three.  This generalizes the result found in~\cite{FASO2} for
vacuum spacetimes.  Here, when the proportional factor $\zeta$
is constant, one can see that this leads to the additional condition
$\rho = 0$, which means that we are within the Kundt class of 
spacetimes~\cite{KUNDT}.

%%%%%%%%%%%%%%%%%%%%%%%%%%%%%%%%%%%%%%%%%%%%%%%%%%%%%%%%%%%%%%%%%%%%%%
%                                                                    %
%                   REMARKS AND CONCLUSIONS                          %
%                                                                    %
%%%%%%%%%%%%%%%%%%%%%%%%%%%%%%%%%%%%%%%%%%%%%%%%%%%%%%%%%%%%%%%%%%%%%%

\section{Remarks and discussion\label{reco}}

In this paper we have discussed the consequences of a symmetry
on the local structure of the spacetime.  We have adopted the
point of view introduced in~\cite{FASO1,FASO2}, where the 
Papapetrou field plays a central role.  We have also generalized
the formalism introduced in~\cite{FASO2} for vacuum spacetimes
to the general case.  In our discussion we have shown that the 
advantages of this type of formalism lie on the consideration
of the following two essential points:  first, that the spacetimes
of General Relativity are four-dimensional, which we have shown
leads to an expression for the components of the Weyl tensor
through the integrability conditions for the components of the
KVF.  In this line, the formalism we have presented is 
based on the Newman-Penrose formalism, in which the fact
that the spacetime is four-dimensional is present.
The second important point is that we have written the NP 
equations in a basis adapted to the algebraic structure of the 
Papapetrou field, and we have added variables, with their 
corresponding equations, that describe the KVF as well as its 
associated Papapetrou field.  By one hand this simplifies the equations,
specially if we want to study cases in which there are
alignments between the principal directions of the Weyl
tensor and the Papapetrou field (see~\cite{FASO2,STEELE1}).
On the other hand, these new variables provide a new
point of view to study the integrability conditions.  
Finally, using the Papapetrou field we introduce 
information about the particular characteristics of the symmetry,
for instance, when we decide whether a KVF has a regular 
or a singular Papapetrou field.

We have illustrated the formalism in the case of spacetimes
with a gravitating electromagnetic field, where we have described 
how the formalism should be applied and we have produced some new results.
The fact that this formalism can be applied to any spacetime with 
a Killing symmetry means that we can use it with a number of
interesting physical situations, and in this sense it could be
used to complement or in combination with other methods of study,
like perturbative schemes or numerical computations of spacetimes.

Finally, it is important to remark that here we have 
dealt only with the case of Killing symmetries however,
the structure of the formalism and the way in which it has been set
up, can be easily translated to other kinds of symmetries.  Actually,
there are some other works in which this has already been done.  
The case of homotheties in vacuum spacetimes has been studied 
in~\cite{STEELE2}, and in~\cite{LUDWIG} it was applied to conformal 
Killing symmetries.

%%%%%%%%%%%%%%%%%%%%%%%%%%%%%%%%%%%%%%%%%%%%%%%%%%%%%%%%%%%%%%%%%%%%%%
%                                                                    %
%                  ACKNOWLEDGEMENTS                                  %
%                                                                    %
%%%%%%%%%%%%%%%%%%%%%%%%%%%%%%%%%%%%%%%%%%%%%%%%%%%%%%%%%%%%%%%%%%%%%%

\ack

Some of the calculations involving the NP formalism were done using the
computer algebraic systems REDUCE and Maple.  We thank the referees
for comments and suggestions that have improved the paper.  F.F. 
acknowledges financial support from the D.G.R. of the Generalitat de 
Catalunya (grant 2000SGR/23), and the Spanish Ministry of Education 
(contract 2000-0606).  C.F.S. is supported by the E.P.S.R.C.

%%%%%%%%%%%%%%%%%%%%%%%%%%%%%%%%%%%%%%%%%%%%%%%%%%%%%%%%%%%%%%%%%%%%%%
%                                                                    %
%                  APPENDICES                                        %
%                                                                    %
%%%%%%%%%%%%%%%%%%%%%%%%%%%%%%%%%%%%%%%%%%%%%%%%%%%%%%%%%%%%%%%%%%%%%%

\appendix

\section{Formalism for spacetimes with a Killing symmetry. NP equations
\label{appa}}

In this Appendix we give the equations of the general framework 
present above using the Newman-Penrose formalism.

\subsection{Equations for the components of the KVF\label{appa1}}

The equations for the components of the KVF [see 
Eqs.~(\ref{ckvf},\ref{kvfnp})]
are obtained by projecting Eq.~(\ref{defpap}) onto 
an adapted NP basis.  The result is the following (they are the same
as those found in~\cite{FASO2} for the vacuum case):
\begin{eqnarray}
&\,&D\xi_k-(\epsilon+\bar{\epsilon})\xi_k+\bar{\kappa}\xi_m
+\kappa\bar{\xi}_m = 0 \,, \label{ckvf1}                          \\
&\,&\triangle\xi_k-(\gamma+\bar{\gamma})\xi_k+\bar{\tau}\xi_m
+\tau\bar{\xi}_m = \textstyle{1\over2}\al \,,            \\
&\,&\delta\xi_k-(\bar{\alpha}+\beta)\xi_k+\bar{\rho}\xi_m
+\sigma\bar{\xi}_m = 0\,,                                         \\
&\,&D\xi_l+(\epsilon+\bar{\epsilon})\xi_l-\pi\xi_m
-\bar{\pi}\bar{\xi}_m = -\textstyle{1\over2}\al \,,      \\
&\,&\triangle\xi_l+(\gamma+\bar{\gamma})\xi_l-\nu\xi_m
-\bar{\nu}\bar{\xi}_m = 0\,,                                      \\
&\,&\delta\xi_l+(\bar{\alpha}+\beta)\xi_l-\mu\xi_m
-\bar{\lambda}\bar{\xi}_m = \textstyle{1\over2} \bar{\theta}\,,     \\
&\,&D\xi_m-(\epsilon-\bar{\epsilon})\xi_m
-\bar{\pi}\xi_k +\kappa\xi_l = 0\,,                               \\
&\,&\triangle\xi_m-(\gamma-\bar{\gamma})\xi_m
-\bar{\nu}\xi_k +\tau\xi_l = -\textstyle{1\over2} \bar{\theta}\,,   \\
&\,&\delta\xi_m+(\bar{\alpha}-\beta)\xi_m
-\bar{\lambda}\xi_k +\sigma\xi_l = 0 \,,                          \\
&\,&\bar{\delta}\xi_m-(\alpha-\bar{\beta})\xi_m-\bar{\mu}\xi_k
+\rho\xi_l = -\textstyle{1\over2} i\bb \,,   \label{ckvf10}       
\end{eqnarray}
where $(D,\triangle,\delta,\bar{\delta})$ are the directional 
derivatives along the NP basis vectors
\[ D = k^a\partial_a \,, ~~\triangle = \ell^a\partial_a\,,~~
\delta = m^a\partial_a\,,~~
\bar{\delta} = \bar{m}{}^a\partial_a\,.\] 
These equations include at the same time both the regular and 
singular cases.  The regular case is obtained by taking $\theta = 0$
and for the singular case we need to take $\al = \bb = 0\,.$

\subsection{Equations for the components of the Papapetrou field\label{appa2}}

They are the projection of the Maxwell equations~(\ref{maxe}) onto
the adapted NP basis.  In the regular case they are equations for 
the complex quantities \al\plusi\bb, actually equations~(\ref{phic}) :
\begin{eqnarray}
\fl D(\al\plusi\bb) & = &  2\rho(\al\plusi\bb) - 4\left\{ 
\xi_k(\Phi_{11}-3\Lambda) + \xi_l\Phi_{00}-\xi_m \bar{\Phi}_{01} -
\bar{\xi}_m \Phi_{01}\right\} \,, \label{maxr1}  \\
\fl \triangle(\al\plusi\bb) & = & - 2 \mu (\al\plusi\bb) + 
4\left\{ \xi_k\Phi_{22} +\xi_l(\Phi_{11}-3\Lambda) - \xi_m\bar{\Phi}_{12} 
- \bar{\xi}_m \Phi_{12}\right\} \,,  \\
\fl \delta(\al\plusi\bb) & = &  2 \tau (\al\plusi\bb) - 4\left\{ 
\xi_k \Phi_{12} + \xi_l \Phi_{01} -\xi_m(\Phi_{11}+3\Lambda) 
-\bar{\xi}_m\Phi_{02}\right\} \,,  \\
\fl \bar{\delta}(\al\plusi\bb) & = & -2 \pi(\al\plusi\bb) + 
4\left\{\xi_k\bar{\Phi}_{12}+\xi_l\bar{\Phi}_{01}-\xi_m \bar{\Phi}_{02} 
- \bar{\xi}_m(\Phi_{11}+3\Lambda)\right\} \,. \label{maxr4}
\end{eqnarray}
Then, we have an expression for all the directional
derivatives of \al~and \bb. In the singular case, Maxwell equations reduce
to equations~(\ref{varphic}) for $\theta$, which once project onto the 
adapted NP basis are
\begin{eqnarray}
D\theta & = & (\rho-2\epsilon)\theta + 2\left\{ \xi_k\bar{\Phi}_{12} 
+\xi_l\bar{\Phi}_{01}-\xi_m\bar{\Phi}_{02}-\bar{\xi}_m(\Phi_{11} 
+3\Lambda)\right\} \,, \label{maxs1}                                 \\
\delta\theta & = & (\tau-2\beta)\theta + 2\left\{ \xi_k\Phi_{22} 
+\xi_l(\Phi_{11}-3\Lambda)-\xi_m\bar{\Phi}_{12} 
-\bar{\xi}_m\Phi_{12} \right\} \,,                                   \\
\textstyle{1\over2}\kappa \theta & = & \xi_k(\Phi_{11}-3\Lambda) 
+\xi_l \Phi_{00} -\xi_m\bar{\Phi}_{01}-\bar{\xi}_m\Phi_{01}\,,       \\
\textstyle{1\over2}\sigma\theta & = & \xi_k\Phi_{12}  
+\xi_l\Phi_{01} - \xi_m(\Phi_{11}+3\Lambda)-\bar{\xi}_m\Phi_{02} 
\,. \label{maxs4} 
\end{eqnarray}
In this case, we only get two directional derivatives of $\theta$,
in the directions of $\mb{k}$ and $\mb{m}$.  The other two 
equations are algebraic relations, from where we can recover
the Goldberg-Sachs and Mariot-Robinson theorems~\cite{GOSA,MARI,ROBI} 
in the particular case of spacetimes with a KVF.

\subsection{Integrability conditions for the components of
the KVF\label{appa3}} 

The integrability conditions for the components of $\mb{\xi}$
lead to an expression for the Weyl tensor [Eqs.~(\ref{defa},
\ref{cccc})].  From it we can compute the Weyl complex scalars
$\Psi_A\,.$  In the regular case, we get the following
expressions: 

\begin{eqnarray}
\fl \Psi_0 = \frac{\al\plusi\bb}{N}(\kappa\xi_m-
\sigma\xi_k)-\frac{2}{N}(\xi_m^2\Phi_{00}-2\xi_k\xi_m\Phi_{01}
+\xi_k^2\Phi_{02}) \,, \label{rpsi0}                             \\
\fl \Psi_1 = \frac{\al\plusi\bb}{N}(\kappa\xi_l
-\sigma\bar{\xi}_m)-\frac{2}{N}\left\{\xi_l\xi_m\Phi_{00} 
-(\xi_k\xi_l+\xi_m\bar{\xi}_m)\Phi_{01}+\xi_k\bar{\xi}_m
\Phi_{02}\right\} \,,                                            \\
\fl \Psi_2 = \frac{\al\plusi\bb}{N}(\rho\xi_l
-\tau\bar{\xi}_m)-\frac{2}{N}\left\{\xi_l^2\Phi_{00}-
2\xi_l\bar{\xi}_m\Phi_{01}+\bar{\xi}_m^{\,2}\Phi_{02}\right\}
-2\Lambda \,,                                   \\
\fl \Psi_3 = \frac{\al\plusi\bb}{N}(\nu\xi_k
-\lambda\xi_m)-\frac{2}{N}\left\{\xi_l\xi_m\bar{\Phi}_{02}
-(\xi_k\xi_l+\xi_m\bar{\xi}_m)\bar{\Phi}_{12}+\xi_k\bar{\xi}_m
\Phi_{22}\right\} \,,                                            \\
\fl \Psi_4 = \frac{\al\plusi\bb}{N}
(\nu\bar{\xi}_m-\lambda\xi_l)-\frac{2}{N}(\xi_l^2
\bar{\Phi}_{02}-2\xi_l\bar{\xi}_m\bar{\Phi}_{12}+
\bar{\xi}_m^{\,2}\Phi_{22}) \,. \label{rpsi4}
\end{eqnarray}
Here we have used that $N = -2\xi_k\xi_l+2\xi_m\bar{\xi}_m$.
In the singular case, the complex scalars $\Psi_A$ can be written
as follows: 
\begin{eqnarray}
\fl \Psi_0 = -\frac{2}{N}(\xi_m^2\Phi_{00}-2\xi_k\xi_m\Phi_{01}
+\xi_k^2\Phi_{02}) \,, \label{spsi0}                     \\
\fl \Psi_1 = -\frac{2}{N}\left\{\xi_l\xi_m\Phi_{00} 
-(\xi_k\xi_l+\xi_m\bar{\xi}_m)\Phi_{01}+\xi_k\bar{\xi}_m
\Phi_{02}\right\} \,,                                     \\
\fl \Psi_2 = -\frac{2}{N}\left\{\xi_l^2\Phi_{00}-
2\xi_l\bar{\xi}_m\Phi_{01}+\bar{\xi}_m^{\,2}\Phi_{02}\right\}
-2\Lambda \,,                                                \\
\fl \Psi_3 = -\frac{1}{N}(\rho\xi_l-\tau\bar{\xi}_m)\theta
-\frac{2}{N}(\xi^2_l\bar{\Phi}_{01} -2\xi_l\bar{\xi}_m\Phi_{11}
+\bar{\xi}_m^{\,2}\Phi_{12}) \,,                             \\
\fl \Psi_4 = \frac{1}{N}\left\{\bar{\xi}_m(\triangle+2\gamma)-
\xi_l(\bar{\delta}+2\alpha)\right\}\theta -\frac{2}{N}(\xi^2_l
\bar{\Phi}_{02}-2\xi_l\bar{\xi}_m\bar{\Phi}_{12}+\bar{\xi}_m^{\,2}
\Phi_{22})\,.  \label{spsi4}
\end{eqnarray}

\subsection{Invariance of $F_{ab}$}

From the Ricci identities we get the equations for the Papapetrou field,
the components of the Weyl tensor and some extra relations,
equations~(\ref{extr1}-\ref{extr3}) in the regular case and
equations~(\ref{exts1},\ref{exts2}) in the singular case.
These relations come from the invariance of $F_{ab}$ [Eq.~(\ref{inva})].
In the NP formalism, the equations for the regular case are:
\begin{eqnarray}
\fl \nu\xi_k + \pi\xi_l - \lambda\xi_m - \mu\bar{\xi}_m = 0\,, \label{invr1} \\
\fl \tau\xi_k - \rho\xi_m + \kappa\xi_l - \sigma\bar{\xi}_m = 0\,, \label{invr2}  \\
\fl \mu\xi_k - \rho\xi_l - \pi\xi_m + \tau\bar{\xi}_m = \nonumber \\
\frac{2}{\al\plusi\bb}(\xi_k^2\Phi_{22} - \xi_l^2\Phi_{00}
-2\xi_k\xi_m\bar{\Phi}_{12}+2\xi_l\bar{\xi}_m\Phi_{01}
+\xi_m^2\Phi_{20}-\bar{\xi}_m^2\Phi_{02})\,. \label{invr3}
\end{eqnarray}
And for the singular case:
\begin{eqnarray}
\fl \tau \xi_k- \rho \xi_m +\kappa \xi_l - \sigma \bar{\xi}_m = 0\,,
\label{invs1} \\
\fl \xi_k(\triangle+2\gamma)\theta - \xi_m(\bar{\delta}+2\alpha)\theta+
(\rho\xi_l-\tau\bar{\xi}_m)\theta = \nonumber \\
\fl~~~~~ -2\xi^2_l\bar{\Phi}_{01} +2\xi_l\xi_m\bar{\Phi}_{02}
-2(\xi_k\xi_l+\xi_m\bar{\xi}_m)\bar{\Phi}_{12}
+4\xi_l\bar{\xi}_m\Phi_{11} - 2\bar{\xi}_m^{\,2}\Phi_{12}
+2\xi_k\bar{\xi}_m\Phi_{22} \,. \label{invs2}
\end{eqnarray}

\subsection{NP form of the equations $\mb{dh}=0$}
%{Integrability conditions for the Maxwell equations}

Within the regular case, a particular situation of interest
corresponds to the subcase defined by the condition $\mb{dh} = 0$.
These equations can written in the NP formalism in the following
form:
\begin{eqnarray}
(D+ \epsilon+ \bar{\epsilon})\mu +(\triangle-\gamma- \bar{\gamma})\rho
-\pi \bar{\pi} +\tau \bar{\tau} = 0 \,, \label{intmax1} \\
(\delta +\bar{\pi}- \bar{\alpha} -\beta)\rho - (D -\bar{\rho} -\epsilon
+\bar{\epsilon})\tau+\mu\kappa- \pi\sigma = 0 \,, \\
(\bar{\delta}-\alpha-\bar{\beta})\rho + (D+\epsilon-\bar{\epsilon})\pi
+\mu\bar{\kappa}+ \tau\bar{\sigma} = 0 \,,  \\
(\delta+\bar{\alpha}+\beta)\mu + (\triangle-\gamma+\bar{\gamma})\tau 
-\rho\bar{\nu}-\pi\bar{\lambda}
= 0 \,, \\
(\triangle + \bar{\mu}+\gamma-\bar{\gamma}) \pi- (\bar{\delta}-
\bar{\tau}+ \alpha+ \bar{\beta})\mu -\tau\lambda+\rho\nu = 0 \,.
\label{intmax5}
\end{eqnarray}

%%%%%%%%%%%%%%%%%%%%%%%%%%%%%%%%%%%%%%%%%%%%%%%%%%%%%%%%%%%%%%%%%%%%%%
%                                                                    %
%                  REFERENCES                                        %
%                                                                    %
%%%%%%%%%%%%%%%%%%%%%%%%%%%%%%%%%%%%%%%%%%%%%%%%%%%%%%%%%%%%%%%%%%%%%%


\begin{thebibliography}{99}
\bibitem{NOTE} Note the misprint in equation~(7.63) of~\cite{KSHM}: instead of
``$\ldots-\triangle\Psi_4\ldots$'' it should read ``$\ldots-D\Psi_4\ldots\,.$''
\end{thebibliography}
\end{document}